# Anomalously low dielectric constant of confined water

L. Fumagalli[1,2]*, A. Esfandiar[3], R. Fabregas[4,5], S. Hu[1,2], P. Ares[1,2], A. Janardanan[1,2], Q. Yang[1,2], B. Radha[1,2], T. Taniguchi[6], K. Watanabe[6], G. Gomila[4,5], K. S. Novoselov[1,2], A. K. Geim[1,2]*

[1] *School of Physics & Astronomy, University of Manchester, Manchester M13 9PL, UK.*
[2] *National Graphene Institute, University of Manchester, Manchester M13 9PL, UK.*
[3] *Department of Physics, Sharif University of Technology, P.O. Box 11155-9161, Tehran, Iran.*
[4] *Departament d'Electrònica, Universitat de Barcelona, C/ Martí i Franquès 1, 08028 Barcelona, Spain.*
[5] *Institut de Bioenginyeria de Catalunya, C/ Baldiri i Reixac 15-21, 08028 Barcelona, Spain.*
[6] *National Institute for Materials Science, 1-1 Namiki, Tsukuba 305-0044, Japan.*

\* Email: *laura.fumagalli@manchester.ac.uk; geim@manchester.ac.uk*

**The dielectric constant $\varepsilon$ of interfacial water has been predicted to be smaller than that of bulk water ($\varepsilon \approx 80$) because the rotational freedom of water dipoles is expected to decrease near surfaces, yet experimental evidence is lacking. We report local capacitance measurements for water confined between two atomically-flat walls separated by various distances down to 1 nm. Our experiments reveal the presence of an interfacial layer with vanishingly small polarization such that its out-of-plane $\varepsilon$ is only ~ 2. The electrically dead layer is found to be two to three molecules thick. These results provide much needed feedback for theories describing water-mediated surface interactions and behavior of interfacial water, and show a way to investigate the dielectric properties of other fluids and solids under extreme confinement.**

Electric polarizability of interfacial water determines the strength of water-mediated intermolecular forces, which in turn impacts a variety of phenomena including surface hydration, ion solvation, molecular transport through nanopores, chemical reactions and macromolecular assembly, to name but a few[1-3]. The dielectric properties of interfacial water have therefore attracted intense interest for many decades[4-7] and, yet, no clear understanding has been reached[8-11]. Theoretical[12-14] and experimental studies[15-17] have shown that water exhibits layered structuring near surfaces, suggesting that it may form ordered (ice-like) phases under ambient conditions. Such ordered water is generally expected to exhibit small polarizability because of surface-induced alignment of water molecular dipoles which are then difficult to reorient by applying an electric field[7-10]. Despite a massive amount of literature dedicated to the subject (see, for example, refs [4-11]), the dielectric constant of interfacial water and its depth remain essentially unknown because measurements are challenging.

The previous experiments to assess $\varepsilon$ of interfacial water mostly relied on broadband dielectric spectroscopy applied to large-scale naturally-occurring systems such as nanoporous crystals, zeolite powders and dispersions[4,5,10,18,19]. These systems allow sufficient amount of interfacial water for carrying out capacitance measurements but the involved complex geometries require adjustable parameters and extensive modelling, which result in large and poorly controlled experimental uncertainties. For example, the extracted values of $\varepsilon$ are strongly dependent on assumptions about the interfacial layer thickness. For the lack of direct probes to measure the polarizability of interfacial water, most evidence has come so far from molecular dynamics (MD) simulations, which also involve certain assumptions. These studies generally predict that the polarizability should be reduced by approximately an order of magnitude[7-9] but the quantitative accuracy of these predictions is unclear because the same simulation approach struggles to reproduce the known $\varepsilon$ for bulk water phases[20]. In this work, we used slit-like channels of various heights $h$ which could be controllably filled with water. The channels were incorporated into a capacitance circuit with exceptionally high sensitivity to local changes in dielectric properties, which allowed us to determine the out-of-plane dielectric constant $\varepsilon_\perp$ of the water confined inside.

The studied devices were fabricated by van der Waals assembly[21] using three atomically flat crystals of graphite and hexagonal boron nitride (hBN) following a recipe reported previously[22,23] (see Supplementary Method S1 and Fig. S1). Here graphite serves as a bottom layer for the assembly as well as the ground electrode in capacitance measurements (Fig. 1a). Next a spacer layer was placed on top of graphite. It was an hBN crystal patterned into parallel stripes. The assembly was completed by placing another hBN crystal on top (Figs 1b,c). The spacer determined the channels' height $h$, and the other two crystals served as top and bottom walls. The reported channels were usually ~ 200 nm wide and several micrometers long. Each of our devices for a given $h$ contained several channels in parallel (Fig. 1), which ensured high reproducibility of our measurements and reduced statistical errors. When required, the channels could be filled with water through a micrometer-size inlet etched in graphite from the back[22,23] (Fig. 1a).

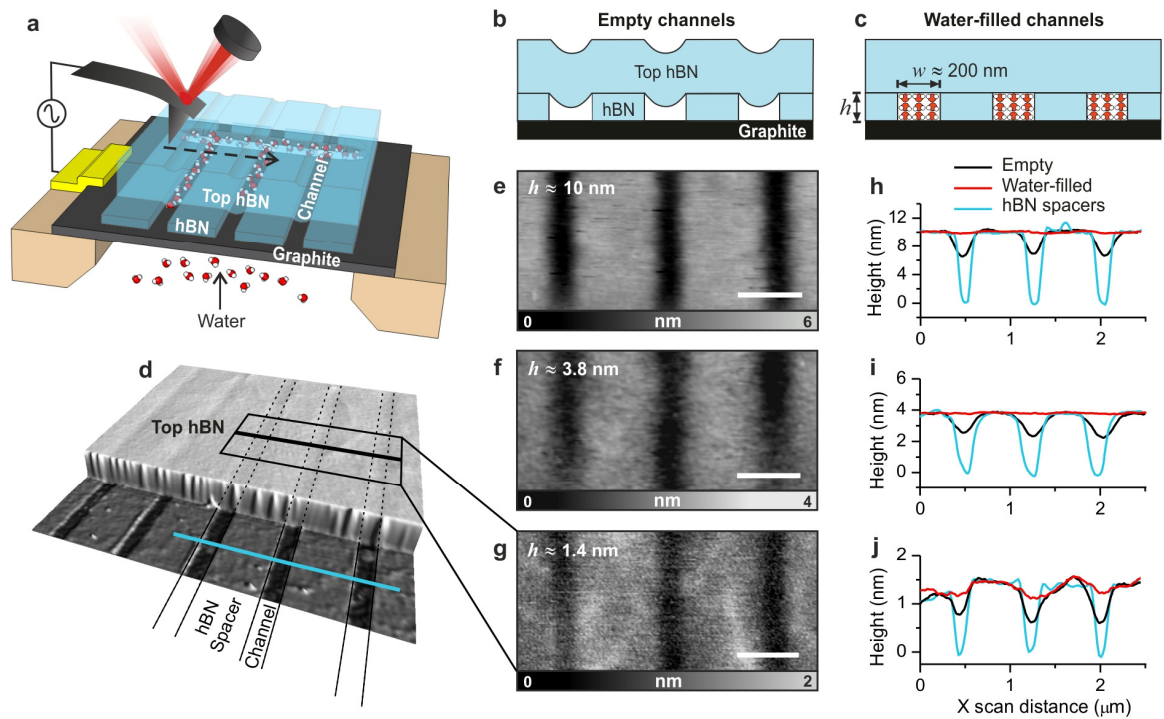

**Fig. 1 | Experimental setup for dielectric imaging. a** Its schematic. The top layer and side walls made of hBN are shown in blue; graphite serving as the ground electrode is in black. The three-layer assembly covers an opening in a silicon nitride membrane (light brown). The channels are filled with water from the back. The AFM tip, kept always in a dry nitrogen atmosphere, served as the top electrode. **b,c** Cross-sectional schematics before (**b**) and after (**c**) filling the channels with water (not to scale). **d** Three-dimensional topography image of one of the devices. **e-g** AFM topography of the sagged top hBN for devices with different $h$ before filling them with water. Scale bars: 500 nm. **h-j** Topography profiles for the top layer (black) and the part not covered by hBN (cyan) as indicated by color-coded lines in (**d**). Red curves: Same devices after filling with water.

To probe $\varepsilon$ of water inside the channels, we employed scanning dielectric microscopy based on electrostatic force detection with an atomic force microscope (AFM), adapting the approach described in ref. [24]. Briefly, by applying a low-frequency ac voltage between the AFM tip and the bottom electrode, we could detect the tip-substrate electrostatic force, which translates into the first derivative of the local capacitance $dC/dz$ in the out-of-plane direction $z$. By raster-scanning the tip, a $dC/dz$ (or "dielectric") image was acquired, from which local dielectric properties could be reconstructed (see Supplementary Methods S3 and S4). Note that the use of hBN is essential for these measurements. First, hBN is highly insulating, which allows the electric field generated by the AFM tip to reach the subsurface water without being screened. It is also highly beneficial to have hBN as the side walls (spacers) because this provides a simple reference for comparison between the dielectric properties of hBN ($\varepsilon_\perp \approx 3.5$)[25] and the nearby water of the same thickness (Fig. 1c). As shown below, the latter arrangement yielded a clear dielectric contrast proving that the dielectric constant of confined water strongly changes with decreasing $h$, independently of the modelling.

Unlike the previous reports[22,23], we chose to use relatively thin (30-80 nm) top crystals. This not only allowed us to reach closer to the subsurface water but also to control that the channels were fully filled during the capacitance measurements (see below) (Supplementary Method S2 and Fig. S2). If there was no water inside, the top hBN exhibited notable sagging[22] as illustrated in Fig. 1b. Figures 1e-g show AFM topographic images for representative devices with $h \approx 10$, 3.8 and 1.4 nm under dry conditions. All of them exhibit some sagging, and its extent depends on thickness of the top hBN[22] (black curves in Figs 1h-j). The channel heights $h$ could also be determined from the same images using the areas that were not covered by the top hBN layer (cyan curves). Such initial imaging as well as dielectric imaging after filling the channels was carried out at room temperature and the whole AFM chamber was filled with dry nitrogen.

Figures 2a-c show AFM topographic images for the same three devices and the same scan areas as in Figs 1e-g but after filling the channels with water, which was done by exposing the backside of our devices to deionized water[22] (Fig. 1a). As the channels became filled through the inlet in the bottom graphite, this lessened adhesion between the side and top walls and, consequently, the sagging

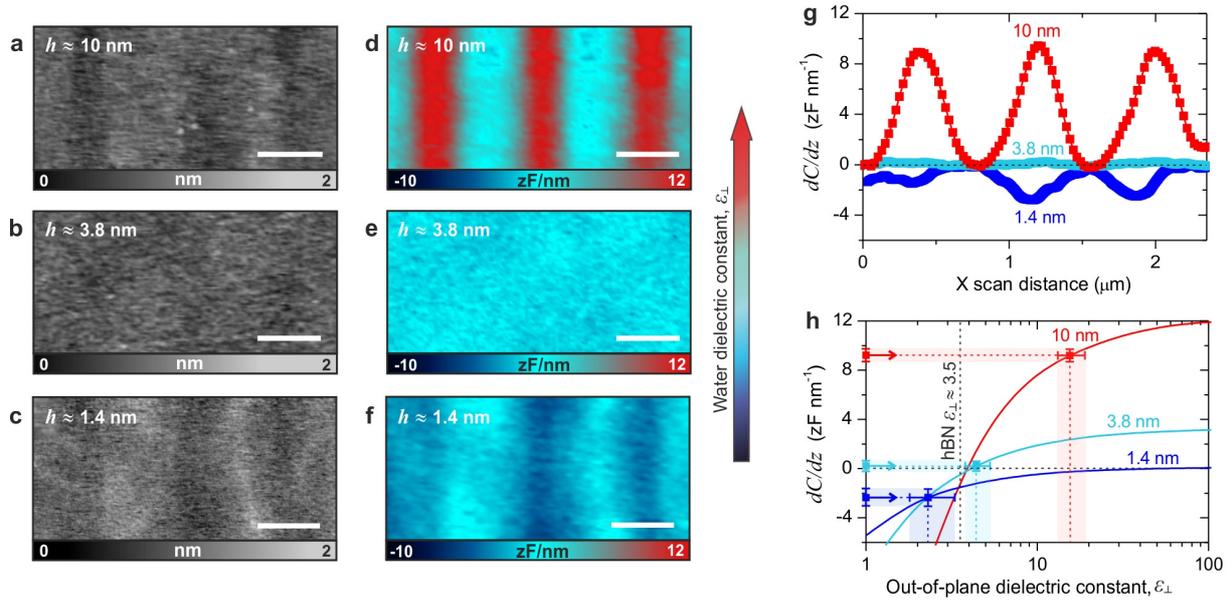

**Fig. 2 | Dielectric imaging of confined water. a-c** Topographic images of the three devices in Fig. 1 after filling them with water. Scale bars: 500 nm. **d-f** Corresponding *dC/dz*. The shown images were obtained by applying a tip voltage of 4 V at 1 kHz (other voltages and frequencies down to 300 Hz yielded similar images). Commercial cantilevers with tips of 100-200 nm in radius were used to maximize the imaging sensitivity. **g** Averaged dielectric profiles across the channels in (**d-f**). **h** Simulated *dC/dz* curves as a function of $\varepsilon_\perp$ for the known geometries of the three shown devices (Shown are the peak values in the middle of the channels). Symbols are the measured values of *dC/dz* from (**g**). Their positions along the *x*-axis are adjusted to match the calculated curves. Bars and light-shaded regions: Standard errors as defined in Supplementary Information.

diminished (Fig. 1c). The top hBN covering water-filled channels became practically straight with little topographic contrast left, independently of *h* (red curves in Figs 1h-j). The corresponding dielectric images for the discussed devices after their filling are shown in Figs 2d-f. One can see very strong contrast which moreover reverses with *h*. For the case of the 10 nm-channels, the red regions containing subsurface water indicate $\varepsilon_\perp$ larger than that of hBN, as expected (Figs 2d and 2g, red). On the contrary, for the 3.8 nm-thick water, the dielectric contrast practically disappeared (Figs 2e and 2g, cyan) whereas the 1.4 nm-thick water exhibited the opposite, negative contrast (Figs 2f and 2g, blue). The images show that the polarizability of confined water strongly depends on its thickness *h* and can reach values smaller than that of hBN with its already modest $\varepsilon_\perp \approx 3.5$. As mentioned above, a reduction in $\varepsilon_\perp$ for strongly confined water is generally expected on the basis of atomistic simulations[7-9] but the observed decrease is much stronger than predicted ($\varepsilon_\perp \approx 10$) or commonly assumed in the literature.

To quantify the measured local capacitance and find $\varepsilon_\perp$ for different water thicknesses, we use a three-dimensional electrostatic model that takes into account the specific geometry of the measured devices as well as of the used AFM tips (see Supplementary Method S5 and Figs S3, S4). The model allows numerical calculation of *dC/dz* as a function of $\varepsilon_\perp$ for a dielectric material inside the channels.

Figure 2h shows the resulting curves for the discussed three devices in Figs 2a-c. By projecting the measured capacitive signals (symbols on the *y*-axis of Fig. 2h) onto the *x*-axis, we find $\varepsilon_\perp \approx 15.5$, 4.4 and 2.3 for $h \approx 10$, 3.8 and 1.4 nm, respectively. We emphasize that $\varepsilon_\perp$ is the only unknown in our model as all the other parameters were determined experimentally. Also, note that some devices exhibited small (few Å) residual sagging in the filled state (see, e.g., Figs 2a,c). If not taken into account, this effect can lead to systematic albeit small errors in determining $\varepsilon_\perp$ (by effectively shifting the calculated curves in the *y*-direction). Our calculations included this residual sagging, too (Fig. S5).

We repeated such experiments and their analysis for more than 40 devices with *h* ranging from ~ 1 to 300 nm. The results are summarized in Fig. 3 which shows the found $\varepsilon_\perp$ as a function of *h*. The bulk behavior ($\varepsilon_\perp \approx 80$) recovers only for water as thick as ~ 100 nm, showing that the confinement can affect the dielectric properties of even relatively thick water layers (Fig. S6). At smaller thicknesses, $\varepsilon_\perp$ evolves approximately linearly with *h* and approaches a limiting value of ~ 2.1 ± 0.2 at $h < 2$ nm where only a few layers of water can fit inside the channels. Note that the functional dependence in Fig. 3 is independent of varying details of our experimental geometries such as, e.g., thickness of the top hBN layer and the AFM tip radius (Fig. S7).

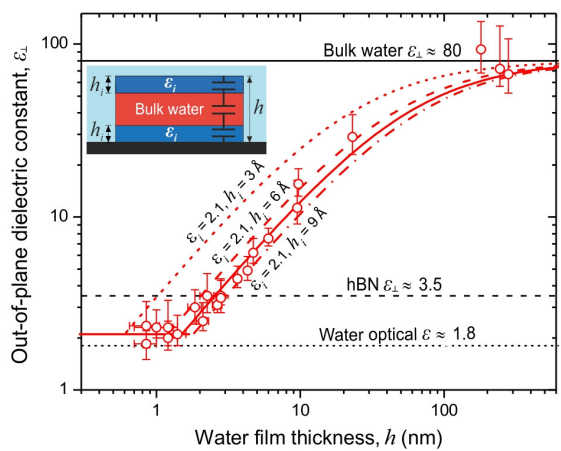

**Fig. 3 | Dielectric constant of water under strong confinement.**
Symbols: $\varepsilon_\perp$ for water channels with different $h$. The *y*-axis error is the uncertainty in $\varepsilon_\perp$ that follows from the analysis such as in Fig. 2h. The *x*-error bars show the uncertainty in the water thickness including the residual sagging. Red curves: Calculated $\varepsilon_\perp(h)$ behavior for the model sketched in the inset. It assumes the presence of near-surface layer with $\varepsilon_i = 2.1$ and thickness $h_i$ whereas the rest of the channel contains the ordinary bulk water. Solid curve: Best fit yielding $h_i = 7.4$ Å. The dotted, dashed and dashed-dotted curves are for $h_i = 3$, 6 and 9 Å, respectively. Horizontal lines: Dielectric constants of bulk water (solid) and hBN (dashed). The dielectric constant of water at optical frequencies (square of its refractive index) is shown by the dotted line.

The dielectric constant $\varepsilon_\perp \approx 2.1$ measured for few-layer water is exceptionally small. Not only it is much smaller than that of bulk water ($\varepsilon \approx 80$) and proton-disordered ice phases such as ordinary ice Ih ($\varepsilon \approx 99$)[26,27] but the value is also smaller than that in low-temperature proton-ordered ices ($\varepsilon \approx 3\text{-}4$)[27]. Moreover, the found $\varepsilon_\perp$ is small even in comparison with the high-frequency dielectric constant $\varepsilon_\infty$ due to dipolar relaxation ($\varepsilon_\infty \approx 4\text{-}6$ for liquid water[28,29] and $\varepsilon_\infty \approx 3.2$ for ice Ih[26,27]). Nonetheless, $\varepsilon_\perp \approx 2.1$ lies – as it should – above $\varepsilon \approx 1.8$ for water at optical frequencies[26,29], which is the contribution due to the electronic polarization. The above comparison implies that the dipole rotational contribution is completely suppressed, at least in the direction perpendicular to the atomic planes of the confining channels. This agrees with the MD simulations that find water dipoles to be oriented preferentially parallel to moderately hydrophobic surfaces such as hBN and graphite[12-14]. The small $\varepsilon_\perp$ also suggests that the hydrogen-bond contribution which accounts for the unusually large $\varepsilon_\infty \approx 4\text{-}6$ in bulk water[28,29] is suppressed, too. The remaining polarizability can be attributed mostly to the electronic contribution (not expected to change under the confinement) plus a small contribution from atomic dipoles, similar to the case of non-associated liquids[29].

Although the found $\varepsilon_\perp$ remains anomalously small (< 20) over a wide range of $h$ up to 20 nm (Fig. 3), this does not actually mean that the polarization suppression extends over the entire volume of the confined water. Indeed, the capacitance response comes from both interfacial and inner molecules, effectively averaging their contributions over the channel thickness. To this end, we recall that water near solid surfaces is believed to have a pronounced layered structure which extends approximately 10 Å into the bulk[12-17]. Accordingly, the found dependence $\varepsilon_\perp(h)$ can be attributed to a cumulative effect from the thin near-surface layer with the low dielectric constant $\varepsilon_i$ whereas the rest of the water has the normal, bulk polarizability, $\varepsilon_{bulk} \approx 80$. The overall effect can be described by three capacitors in series as shown in the inset of Fig. 3. This model yields the effective $\varepsilon_\perp = h/[2h_i/\varepsilon_i + (h - 2h_i)/\varepsilon_{bulk}]$ where $h_i$ is the thickness of the near-surface layer. Its $\varepsilon_i$ can be taken as $\approx 2.1$ in the limit of small $h$ if we assume that the layered structure does not change much with increasing $h$ (ref.[13]) and is similar at both graphite and hBN surfaces, as the MD simulations predict[14]. Figure 3 shows that the proposed simple model describes well the experimental data, allowing an estimate for the thickness $h_i$ of interfacial water with the suppressed polarization (see Supplementary Method S9 and Fig. S8). Within the experimental error, our data yield $h_i \approx 7.5 \pm 1.5$ Å, in agreement with the expected layered structure of water[14-17]. In other words, the electrically dead layer extends two-three molecular diameters away from the surface. This is also consistent with the thickness $h = 1.5\text{-}2$ nm – the double of $h_i$ – where the limiting value $\varepsilon_\perp \approx 2.1$ is reached (see Fig. 3), which can be understood as the distance at which the near-surface layers originating from top and bottom walls begin to merge.

To conclude, we have succeeded in the long-lasting quest to measure the dielectric constant of confined water at the nanoscale. Our results are important for better understanding of long-range interactions in biological systems, including those responsible for the stability of macromolecules such as DNA and proteins, and of the electric double layer that plays a critical role in electrochemistry, energy storage, etc. The results can also be used to fine tune parameters in future atomistic simulations of confined water.

# Supplementary Information

**S1. Device fabrication**

We made our devices following fabrication procedures similar to those reported in refs [22,23]. In brief, a free-standing SiN membrane (500 nm in thickness) was made from a commercial Si/SiN wafer and used as a substrate for the van der Waals assembly (see Fig. S1; purple). A rectangular aperture of ≈ 3×25 $\mu m^2$ in size was then etched in the membrane (Fig. S1). This aperture served later as an inlet to fill the nanochannels with water from a reservoir connected to the back of the wafer (Fig. 1a of the main text). Next, we transferred a large cleaved graphite crystal (thickness of ~ 10-50 nm) to seal the aperture. Separately, an hBN crystal referred to as spacer was prepared on another substrate and patterned into parallel stripes using e-beam lithography and reactive ion etching. The hBN spacer had thickness $h$ chosen in the range of ~ 1-300 nm. The stripes were spaced apart by ~ 200 nm and had widths of 0.5-1.5 μm. The spacer stripes were then transferred onto the bottom graphite and aligned perpendicular to the long-axis of the aperture in the SiN membrane (Fig. S1b). As the next step, reactive ion etching was used again from the back of the Si/SiN wafer to project the aperture onto the hBN-on-graphite assembly. The second hBN crystal referred to as top-hBN was prepared with a thickness of 30-80 nm and transferred on top of the assembly. As a result, we obtained an array of channels with the height $h$ and width ~ 200 nm. The top hBN crystal sealed the etched opening so that the only path from the back side of the SiN membrane to its top was through the resulting nanochannels. After each transfer, we annealed our assembly in Ar/$H_2$ at 400°C for 3 hours to remove any polymer residue and other contamination. Finally, we made an electrical contact to the bottom graphite using photolithography and e-beam evaporation of Au. Optical images of a representative device with the channel height $h ≈ 4$ nm are shown in Fig. S1.

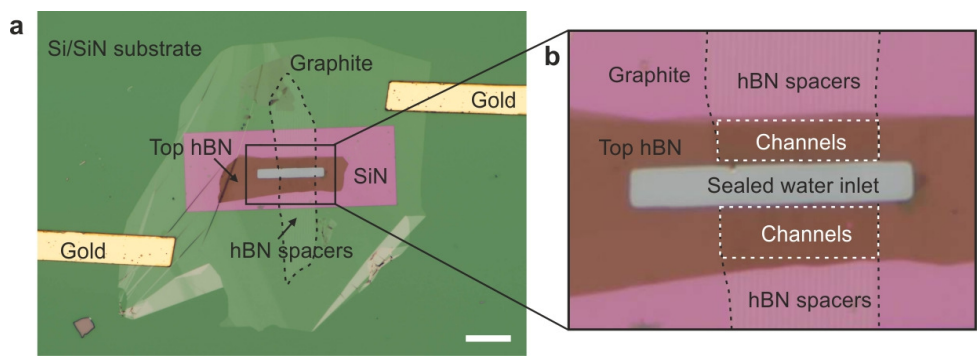

**Fig. S1 | Devices for local dielectric imaging of confined water. a** Optical micrograph of one of our devices. The top hBN layer is ~ 45 nm thick and $h ≈ 4$ nm. The free-standing SiN membrane appears in purple; the Si/SiN wafer in green. The graphite layer is contacted with gold pads to serve as the ground electrode. Scale bar: 10 μm. **b** Zoom into the central region of (**a**). The areas with nanochannels are shown by the two dashed rectangles. Regions with the hBN spacers not covered by the top hBN and used to measure $h$ are outlined by black dashes.

**S2. Filling nanochannels with water**

It was essential to verify that there was water inside nanochannels probed by our scanning probe approach. In particular, we needed to ensure that individual channels under investigation were neither empty nor contained another material because in principle they could be, for example, blocked by contamination or filled with a polymer residue. Global measurements such as those reported previously[22,23], in which a water flow through hundreds of channels was detected, were insufficient for the purpose of our study. Note that we could see the water inside individual channels with $h > 100$ nm using optical microscopy but water was invisible for $h < 20$ nm as illustrated by Fig. S2. Here one can clearly see that water filled all the large channels connected to the water inlet, except for one that is probably blocked by contamination (see Fig. S2a). On the contrary, Fig. S2b shows that the optical contrast was insufficient to detect water inside channels with small $h$ or actually even see such small channels.

To verify the presence of water in the latter case, we adopted the following strategy. The thickness of the top hBN layer was chosen deliberately in the range of typically 30 to 50 nm, which allowed the hBN cover to sag inside the channels if they were empty (in dry air) as sketched in Fig. 1b of the main text. Upon filling them with water from the backside inlet, the top layer straightened (Fig. 1c of the main text). Accordingly, by monitoring topographic changes in the top hBN position before and after (see Figs 1e-g and Figs 2a-c, respectively), we could ensure that individual channels under investigation were first empty and then filled with water for their dielectric imaging. Importantly, topographic and dielectric AFM images could be acquired one after another without perturbing the experimental setup. Note that if the top hBN sagged completely and touched the bottom graphite or if the channels were blocked by contamination, no straightening of the top hBN occurred. Such channels were obviously excluded from our investigation. This monitoring procedure was working well even for devices with $h < 2$ nm, which required Å-scale topographical imaging to detect sagging and straightening (Fig. 1j of the main text). For such channels, we typically used a slightly thicker top hBN (50 to 80 nm) to avoid its excessive sagging.

We studied more than 40 devices in which the top layer was partially sagged as required for monitoring of the water filling. Our success rate was roughly 50% with the rest of the devices being blocked, most probably because of sagging of a very thin (few nm) part of the top hBN, which could be often found near cleaved edges[23]. For unblocked channels that allowed water inside, our dielectric measurements discussed below were highly reproducible. Note that both topographical and dielectric measurements were always carried out at very low (few %) humidity because the water reservoir attached to the back side of the Si/SiN was completely isolated from the AFM chamber whereas a water flux through the channels themselves was so small[22] that it could not possibly change the humidity even locally.

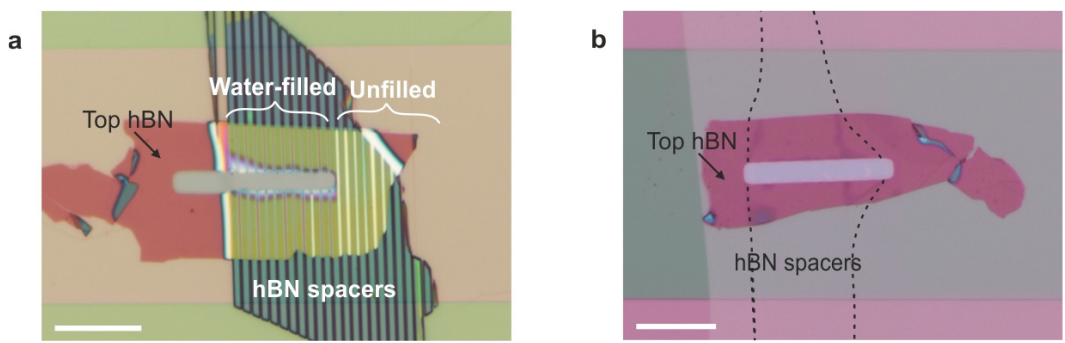

**Fig. S2 | Optical images of our devices after filling them with water. a** Thick device ($h \approx 242$ nm). Channels with water appear darker than the empty channels that are seen to the right of the image and not connected to the inlet (grey rectangle). **b** Thin ($h \approx 3$ nm) device filled with water. Individual channels cannot be resolved on the micrograph. Scale bars: 10 μm.

**S3. Local dielectric imaging**

Dielectric images of the water-filled channels were obtained by a scanning probe technique[24] here referred to as scanning dielectric microscopy. It is based on local electrostatic force detection[30,31] by using an atomic force microscope. Images were taken at room temperature and in a dry atmosphere (relative humidity of few %) using a commercial AFM (*Nanotec Electronica*). After locating a region of interest and taking its topographic image, we scanned the AFM tip at the constant height $z_{scan}$ from the top hBN surface. The dielectric images were acquired at 1 sec per line with an applied ac voltage of typically $V_{ac}$ = 4 V and the frequency $\nu$ = 1,000 Hz, unless stated otherwise. We recorded mechanical oscillations of the AFM cantilever induced by the electrostatic force between the tip and the surface at the double frequency ($2\nu$) using a lock-in amplifier. The first derivative of the tip-substrate capacitance $dC/dz$ in the out-of-plane direction $z$ is given by $dC/dz = D_{2\nu}(z) \cdot 4k/v_{ac}^2$ where $D_{2\nu}$ is the cantilever oscillation amplitude at $2\nu$, and $k$ the spring constant of the cantilever. The expression is valid for frequencies well below the resonance frequency of the cantilever. Images obtained in this mode depend only on the dielectric properties of probed devices, their geometry and the AFM tip geometry. We determined the scan height $z_{scan}$ by recording the tip deflection in the dc mode, and the $dC/dz$ signal was also recorded as a function of the tip-surface distance at image edges, as previously reported[24]. While the deflection-distance curve allowed us to determine $z_{scan}$, we used the $dC/dz$ signal to detect any vertical drift and corrected $z_{scan}$ for it. Typically, $z_{scan}$ was larger than 15 nm to avoid short-range interactions. Note that this approach is different from scanning polarization force microscopy[32] in that the measured force variations at $2\nu$ are not used as a feedback signal in our case. Instead, we turn off the feedback, retract the tip at height $z_{scan}$ from the surface and scan it in a straight line, which minimizes stray capacitance variations and simplifies data analysis. A representative example of dielectric imaging is shown in Fig. S6 for a device with large channels, in which the bulk-water dielectric behavior was recorded.

Before and after taking the dielectric image, we also took $dC/dz$-approach curves over distances of 0-600 nm from the substrate. These curves were used to calibrate the AFM tip geometry *in situ* (see Fig. S7 and refs [24,33-36]). The approach curves were taken directly above top hBN near the scanned area, after verifying that we recovered the same geometrical parameters as measured above the bottom graphite and gold contacts. We used commercial doped-diamond coated probes (CDT-CONTR, *Nanosensors*) with spring constants in the range 0.3 - 1.0 N/m, nominal radii of 100 to 200 nm and the cone half-angle of ~ 30º.

**S4. Dielectric image analysis**

All topographic, dielectric and other AFM data were analyzed using *WSxM*[37] software and custom-made *Matlab* and *Mathcad* routines. To extract the dielectric constant of water, we analyzed changes in $dC/dz$ over filled channels as compared to the derivative measured over hBN spacer regions, that is, not the absolute value of $dC/dz$. For brevity, we below redefine $dC/dz$ as $dC/dz = dC(z_{scan}, \varepsilon_\perp)/dz - dC(z_{scan}, \varepsilon_{hBN})/dz$, where $\varepsilon_{hBN}$ is the out-of-plane dielectric constant of hBN ~ 3.5 (ref. [25]). We then compared the $dC/dz$ detected over the center of the nanochannel (peak value) with the calculated value using our numerical model discussed in the next chapter. $\varepsilon_\perp$ was the only fitting parameter to match the experimental and numerical data. All the necessary geometrical parameters of our samples were experimentally determined using AFM and scanning electron microscopy. We found the geometric parameters of our AFM probes *in situ* by fitting the experimental $dC/dz$ approach curves with their numerical model as described previously[24,33-36]. This allowed us to obtain the effective tip radius $R$ and the cone half-angle $\theta$, which are responsible[24,36] of the local electrostatic interaction, with a high accuracy of ± 3 nm and ± 0.25°, respectively. The spring constants of our cantilevers were given by the manufacturer but we verified that the use of probes with different spring constants did not affect the extracted dielectric constants, in agreement with the result of ref. [24] (see supplementary information therein).

Some dielectric constants $\varepsilon_\perp$ that were found experimentally and are summarized in Fig. 3 of the main text have asymmetric error bars. This is a result of the logarithmic-like dependence of the tip-surface capacitance on $\varepsilon_\perp$ (see the simulated curves in Fig. 2h and Fig. S4e). The feature is typical for local dielectric measurements (see, e.g., ref. [24]) and caused by the use of a sharp tip as the top electrode instead of a planar electrode. The logarithmic dependence is also responsible here for the higher sensitivity of our technique to negative capacitive variations in $dC/dz$ ($\varepsilon_\perp \leq 3.5$) as compared to large positive changes ($\varepsilon_\perp > 10$), as it can be seen in Figs S4b,d,e. This explains why the error bars in Fig. 3 of the main text are large for thick water ($h > 10$ nm), despite the $dC/dz$ signal is large in this case.

The experimental parameters used in the calculations for the three devices of Figs 1-2 of the main text are the following. Sample dimensions in Fig. 1e and Figs 2a,d: $h = 10$ nm, top hBN layer thickness $H = 51$ nm, channel width $w = 200$ nm, hBN spacer width $w_s = 800$ nm. For Fig. 1f and Figs 2b,e: $h = 3.8$ nm, $H = 46$ nm, $w = 170$ nm, $w_s = 800$ nm. For Fig. 1g and Figs 2c,f: $h = 1.4$ nm, $H = 39$ nm, $w = 200$ nm, $w_s = 800$ nm. The dielectric images in Figs 2d-f ($h = 10$, 3.8 and 1.4 nm) were measured at scan heights $z_{scan} = 30$, 25 and 17 nm and with tip radii $R = 165$, 137 and 101 nm (half-angle $\theta = 29.0$, 31.5 and 30.5°), respectively. Note that the observed suppression in $\varepsilon_\perp$ is independent of the scan height and the tip radius. We carefully verified this by taking dielectric images at different scan heights (not shown here but see ref. [24]) and with different AFM probes (see Fig. S7). Only the capacitive contrast ($dC/dz$) changes with $z_{scan}$ and $R$, increasing for smaller scan heights and larger radii[24].

**S5. Finite-element numerical simulations**

Three-dimensional finite-element numerical calculations were implemented using COMSOL Multiphysics 5.2a (AC/DC electrostatic module) linked to *Matlab*. The AFM probe was modelled as a truncated cone with half-angle $\theta$ and height $H_{cone}$ terminated with a tangent hemispherical apex of radius $R$ as shown in Fig. S3a. To reduce computational time, the cone height was reduced to 6 µm (half its nominal value) and the cantilever was modeled as a disk of height $H_{cantilever} = 3$ µm and zero length $L_{cantilever}$, thus omitting the cantilever length. We have checked that these approximations have no impact on the extracted dielectric constants for the geometry analyzed here[36,38]. We simulated the probed heterostructure as three buried nanochannels (Fig. S3a). They were modeled as rectangular parallelepipeds of length $l = 2.5$ µm, height $h$, width $w$, spacing $w_s$ (found experimentally as discussed above) and the out-of-plane dielectric constant $\varepsilon_\perp$. The water channels were surrounded from above by a dielectric matrix with $\varepsilon_{hBN} = 3.5$ of a rectangular shape (length $l = 2.5$ µm, width $W$ and height $H + h$). Note that, for our thickest channels ($h > 100$ nm), we modeled them with trapezoidal rather rectangular cross-sections in order to take into account the ~ 55° angle of the lateral walls, which appeared during etching of thick hBN spacer crystals by reactive ion plasma.

For each device, we numerically solved the Poisson's equation for the specific dimensions of the device and the probe with the nanochannel dielectric constant $\varepsilon_\perp$ as the only varying parameter. We calculated the electrostatic force acting on the probe and, therefore, the capacitance first-derivative $dC/dz$ as a function of $\varepsilon_\perp$ by integrating the built-in Maxwell stress tensor on the probe surface. To avoid size effects related to the simulation box, we used a cylindrical box with infinite lateral extension at the top and lateral boundaries by using the built-in infinite-element transformation. The boundary conditions were set as follows: applied voltage of 1 V at the tip surface; zero voltage at the bottom electrode; zero charge at the top and side boundaries. We validated these simulations against analytical formulas for thin films as previously reported[34,38]. Optimization and numerical noise reduction were carried out to meet the accuracy required here. Note that our simulations involved 3D structures with sizes spanning over more than three orders of magnitude - from the micrometer-sized matrix and probe down to the atomically-thin channels. To this end, a mesh of ~ $10^6$ elements was typically required. An example of the electrostatic potential generated around a representative device is shown in Fig. S3b.

Furthermore, we implemented Matlab routines to simulate the tip scanning at a constant height $z_{scan}$ from the top hBN surface as in the experiments. This allowed us to compute dielectric images $dC(x,y)/dz$ where $(x,y)$ is the in-plane tip position. Examples of the calculated images and corresponding profiles along the *x*-axis are plotted in Fig. S4 for representative devices with $\varepsilon_\perp = 2$ and 80. In addition, we also computed fixed-position "spectroscopic" curves, in which the tip was held fixed over the center of a channel and $dC/dz$ was calculated as a function of $\varepsilon_\perp$ with respect to the value computed over the center of the hBN spacer. We used such spectroscopic curves to fit our experimental data and obtain $\varepsilon_\perp$, as shown in Fig. 2h using the real data and in Fig. S4e for simulated ones.

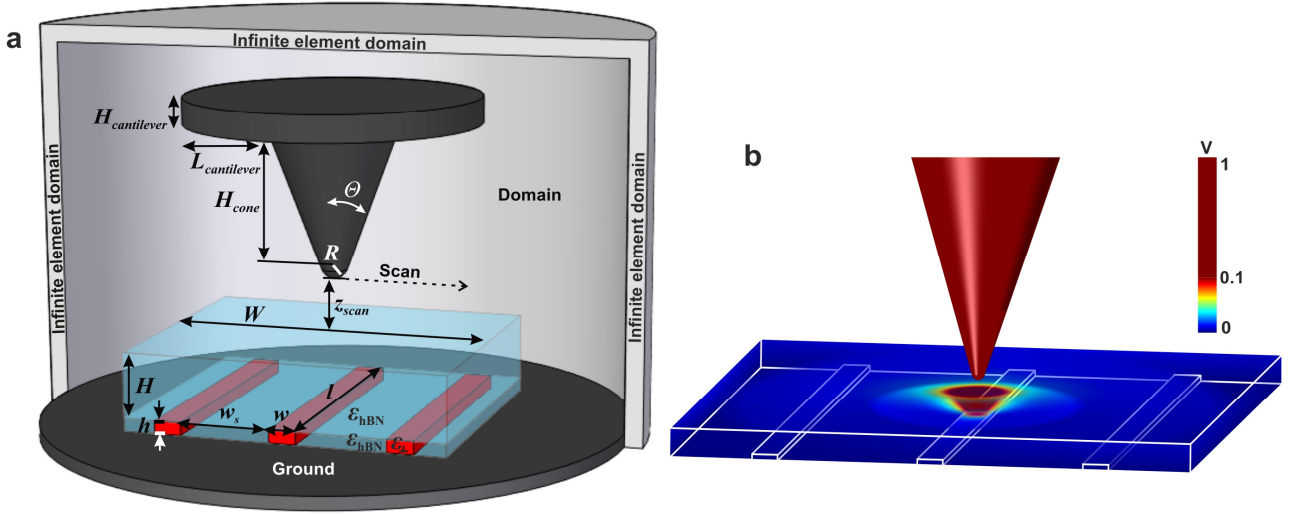

**Fig. S3 | Numerical simulations**. **a** Simplified schematics of our 3D model, including the AFM tip and three nanochannels (not to scale). **b** Example of calculated potential distributions. For clarity, only the potential distribution inside the device is shown. In this case, we used $H = 40$ nm, $h = 10$ nm, $w = 150$ nm, $w_s = 800$ nm; $W = 3$ μm; $\varepsilon_{hBN} = 3.5$ and $\varepsilon_\perp = 2$. AFM tip: $R = 100$ nm, $\theta = 25°$, $H_{cone} = 6$ μm, $H_{cantilever} = 3$ μm, $L_{cantilever} = 0$ μm, $z_{scan} = 20$ nm.

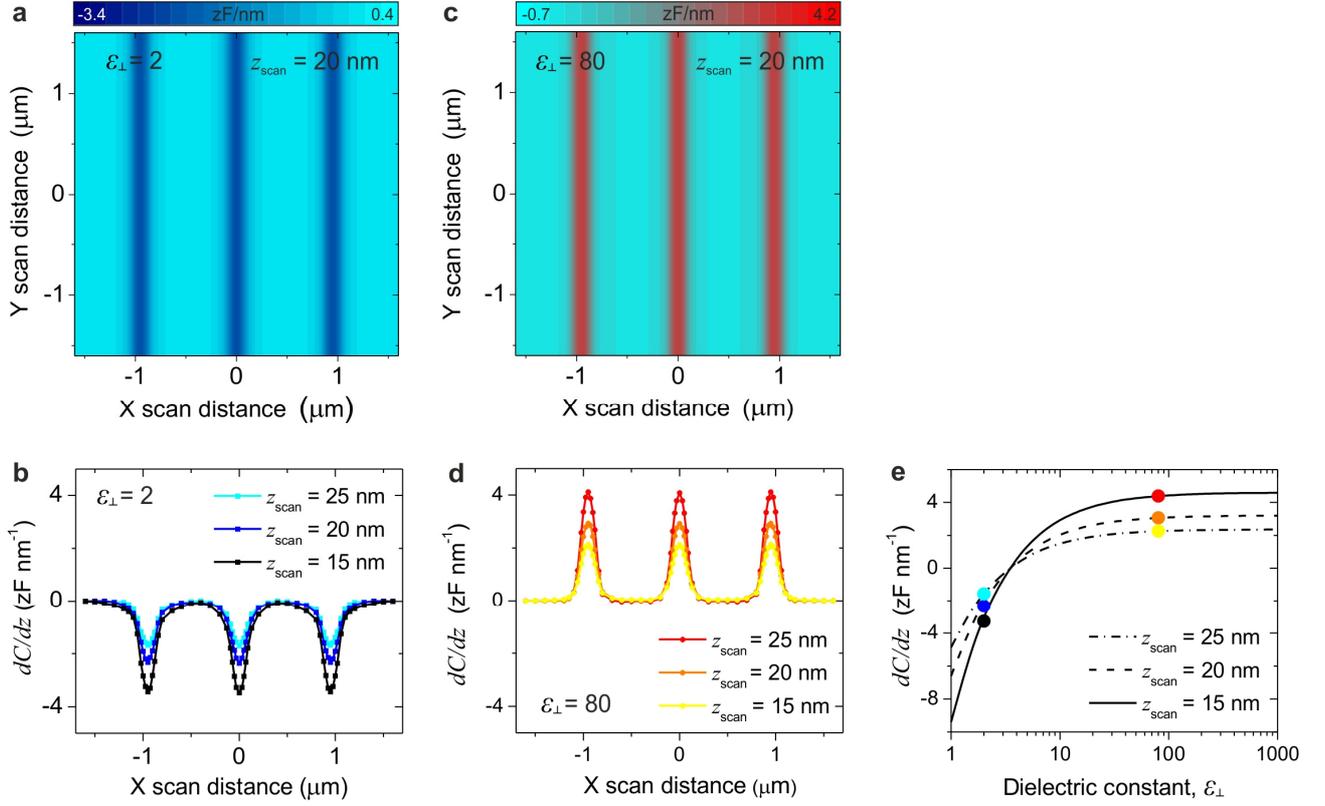

**Fig. S4 | Simulated dielectric images. a,c** Dielectric constant $\varepsilon_\perp = 2$ for (**a**) and $\varepsilon_\perp = 80$ for (**c**). Scan height $z_{scan} = 20$ nm from the top hBN. **b,d** Corresponding profiles for three heights $z_{scan} = 15$, 20 and 25 nm. Relative $dC/dz$ are shown, taken with respect to their values over the hBN spacer. Used parameters: $h = 3$ nm, $H = 40$ nm, $w = 150$ nm, $w_s = 800$ nm, $R = 100$ nm, $\theta = 30°$. **e** Simulated $dC/dz$ as functions of $\varepsilon_\perp$ with the tip fixed at the channel center. Symbols indicate $\varepsilon_\perp = 2$ and 80 for parameters as in (**b**) and (**d**). Note that such "spectroscopic" curves show the contrast inversion at $\varepsilon_\perp = \varepsilon_{hBN} = 3.5$ as well as decrease in sensitivity with increasing $z_{scan}$, as expected.

## S6. Subsurface sensitivity

The ability of electrical scanning probe techniques such as electrostatic force microscopy (EFM), Kelvin probe force microscopy (KPFM) and scanning microwave microscopy (SMM) to obtain subsurface images on the nanoscale is widely acknowledged (see, for example, refs [39-42]). By exploiting the long-range nature of the electrostatic interaction between the tip and a conductive substrate, these techniques are able to detect nanoscale objects buried inside a dielectric matrix. In particular, non-destructive visualization of conductive objects such as carbon nanotubes embedded in dielectrics of hundreds of nm in thickness has previously been reported using EFM[39,40] and KPFM[41]. Our work uses a similar approach based on electrostatic-force detection, which allows detection of water as thin as 1 nm buried ~ 100 nm below. The subsurface sensitivity in our case depends on several parameters. It obviously decreases with the thickness $H$ of the top hBN layer and the scan height. Also, the sensitivity increases with the width and the height of the nanochannels and the AFM tip radius. Accordingly, we used AFM tips with large radii (100-200 nm) rather than probes with small few-nm radii as in ref. [24]. This was intentional to enhance our sensitivity and reach to the water below our relatively thick (40-80 nm) top hBN. The latter thickness was required to avoid the collapse of our nanochannels (see above). Wider channels with $w > 200$ nm would increase sensitivity but, unfortunately, were also nonviable because of the same collapse[22].

## S7. Effect of residual sagging

After filling water inside the studied nanochannels, they often exhibited small residual sagging ($\leq 3$ Å) for $h < 20$ nm, where our thick channel devices ($h > 100$ nm) usually swelled slightly (by 1-2 nm). We verified that these topographic features had no major impact on our results. Moreover, to achieve highest possible accuracy in our experiments, we corrected our numerical modelling by including this residual sagging/swelling for each individual device. The simulation setup is sketched in the inset of Fig. S5a. As an example, Fig. S5a shows the simulated profiles with and without residual sagging by 3 Å for the device of $h = 1.4$ nm and $\varepsilon_\perp = 2$ of Fig. 2 of the main text. In either case the resulting $dC/dz$ variation remains clearly negative, and the profile is slightly higher if the sagging is not included in the model (open symbols). Accordingly, Fig. S5b shows the simulated "spectroscopic" curves used to extract the water's dielectric constant for all three devices of Fig. 2 of the main text. Without including the sagging into our model, the resulting curves for $dC/dz$ would go slightly higher (dashed) than those that take into account the sagging and are shown in Fig. 2h (solid). This would lead to a slight underestimate for the dielectric constants of confined water. Instead of the correct values $\varepsilon_\perp = 15.5, 4.4$ and 2.3 (filled symbols), ignoring the sagging effect ($s = 3, 1.5$ and 3 Å as measured in Fig. 2a-c) could have resulted in $\varepsilon_\perp = 10.2, 3.8$ and 1.65 (open symbols) for $h = 10, 3.8$ and 1.4 nm, respectively. Note that the relative impact of sagging is practically the same for all the three devices independently of their $h$. This behavior can be traced back to the logarithmic decrease in the $dC/dz$ signal with increasing $\varepsilon_\perp$. On one hand, the impact of any small topography artifact is expected to decrease with increasing $h$. One the other hand, this is counterbalanced by the larger uncertainty with increasing water's $\varepsilon_\perp$ due to the logarithmic sensitivity discussed above. Hence, the effect of the residual sagging turns out to be roughly the same for all the channels.

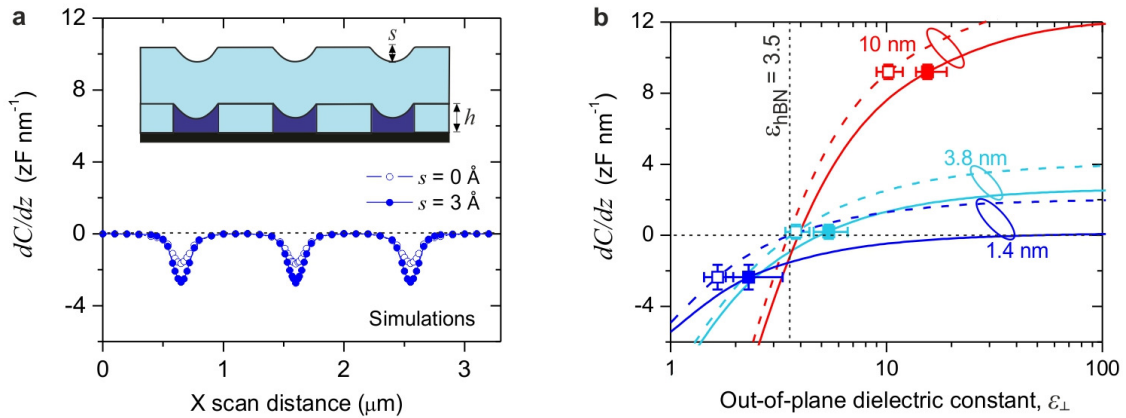

**Fig. S5 | Sagging effect. a** Simulated $dC/dz$ profiles with and without 3 Å-sagging (filled and open symbols, respectively) for the device with $h = 1.4$ nm in Fig. 2 of the main text and $\varepsilon_\perp = 2$. Parameters: $h = 1.1$ nm and the sagging depth $s = 3$ Å (filled symbols); same $h$ and no sagging $s = 0$ (open); the other parameters are as in Fig. 2. Inset: Sketch of the model to include sagging (not to scale). **b** Simulated $dC/dz$ as functions of $\varepsilon_\perp$ with (solid) and without (dashed curves) taking into account the residual sagging for the three specific devices in Fig. 2 of the main text, $s = 3, 1.5$ and 3 Å for $h = 10, 3.8$ and 1.4 nm, respectively, as measured in Figs 2a-c. Device parameters are as in Supplementary Method S4. Open (filled) symbols are the measured $dC/dz$ and their projections onto the $\varepsilon_\perp$ axis without (with) including the sagging.

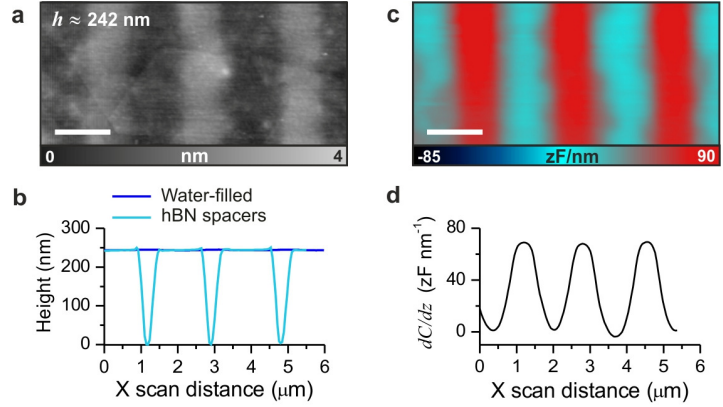

**Fig. S6 | Dielectric imaging of large channels. a** Topographic image and **b** Corresponding profile of a device with $h \approx 242$ nm after filling it with water (blue curve). The topography profile of the hBN spacer is shown in cyan. **c** Corresponding dielectric image and **d** Its averaged profile. Scale bar: 1 μm.

## S8. Effect of the tip radius

We verified that the measured dielectric constant of confined water (Fig. 3 of the main text) was independent of geometry and dimensions of our AFM probes. To this end, we repeated the dielectric measurements using different probes. Their effective tip radii $R$ were measured *in situ* before and after each dielectric imaging experiment because values of $R$ are required in our simulations.

Examples of the approach curves used to extract $R$ are shown in Fig. S7a for the three specific AFM probes employed in the experiments of Fig. 2 of the main text. Fig. S7b plots the measured $R$ of all the probes used in our experiments to extract the dielectric constants in Fig. 3. In this case, $R$ is plotted against the water thickness $h$. One can see that the used $R$ are randomly scattered over the expected range 100-200 nm indicated by the manufacturer and there is no correlation with $h$ or other geometrical parameters of our devices. This shows that the observed reduction in water's $\varepsilon_\perp$ was independent of the used AFM tips. We also confirmed that for the known tip radii and using hBN crystals as test structures, our approach yielded the correct value of $\varepsilon_{hBN} \approx 3.5$ (not shown).

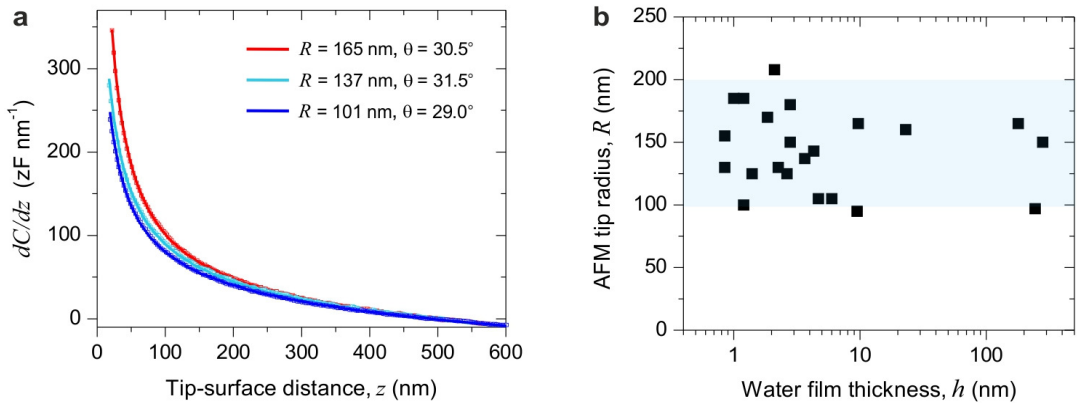

**Fig. S7 | Impact of the tip radius. a** Experimental approach curves (symbols) and their fitting (solid curves) to find tip radii. The data correspond to the experiments of Fig. 2. **b** Measured $R$ for the AFM probes used in our experiments for all $h$ shown in Fig. 3. The blue-shaded region indicates the nominal range expected for these probes.

**S9. Intermediate water thickness**

As described in the main text, the strong suppression of $\varepsilon$ observed for water of intermediate thickness can be readily explained by having a thin near-surface layer that is not polarizable in series with normally polarizable water further into the bulk. The experimental data presented in Fig. 3 of the main text were fitted for $h > 2$ nm using the tri-capacitor model shown in the figure's inset. In the analysis where we used the weighted nonlinear least-squares method, we assumed constant $\varepsilon_\perp = 2.1$ for the interfacial layer (as in our thinnest channels) and $\varepsilon_\perp = 80$ for the bulk water. The best fit yielded the interfacial water thickness $h_i = 7.4$ Å (Fig. 3 and Fig. S8, red solid curve; 95% confidence interval of 7.0–7.8 Å). This estimate agrees well with the widely accepted model of the layered structure of water near surfaces, which extends 2-3 water diameters (~ 3 Å) into the bulk. This model is also consistent with the minimum $\varepsilon_\perp \approx 2.1$ found for $h < 2$ nm.

It is instructive to compare the observed dependence $\varepsilon_\perp(h)$ with that predicted by the above model for other values of the dielectric constant, in particular for those often assumed in the literature. As an example, Fig. S8 shows the calculated curves for $\varepsilon_i = 6$ (roughly one order of magnitude smaller than in bulk water[7-9] and representative of water high-frequency behavior[26-29]). The resulting curves for the interfacial thickness $h_i = 3$, 6 and 9 Å lie well above our experimental results and do not intersect the value $\approx 3.5$ corresponding to hBN's dielectric constant. This illustrates again that it is impossible to explain the obtained dielectric images showing zero and negative contrast without much stronger suppression of $\varepsilon_\perp$ than routinely assumed in the literature for interfacial water.

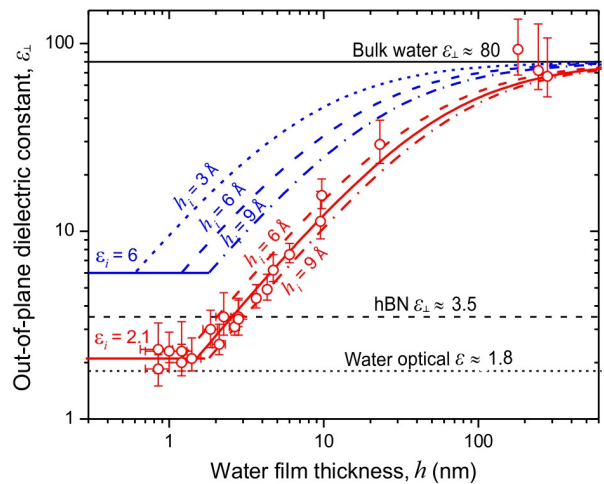

**Fig. S8 | Expected and measured suppression of the dielectric constant in interfacial water.** Data shown in red are same as in Fig. 3 of the main text. Blue curves: $\varepsilon_\perp(h)$ predicted by the same model but using $\varepsilon_i = 6$ instead of 2.1 and thickness $h_i = 3$, 6 and 9 Å (dotted, dashed and dashed-dotted curves, respectively).